\documentclass[preprint,preprintnumbers,nofootinbib,amsmath,amssymb,floatfix]{revtex4-1}

\usepackage[english]{babel}
\usepackage{color}
\usepackage{amsmath}
\usepackage{physics}
\usepackage{xr-hyper}
\usepackage[hidelinks]{hyperref}
\usepackage{diagbox}
\usepackage{multirow}
\usepackage{amsfonts}
\usepackage{amssymb}
\usepackage{latexsym}
\usepackage{graphicx}
\usepackage{adjustbox}
\usepackage{float}
\usepackage{xr}
\usepackage{subcaption}
\usepackage{caption}
\usepackage{enumitem}
\usepackage{threeparttable}
\usepackage{ytableau}
\usepackage[nodisplayskipstretch]{setspace} 

\newcommand{\al}{\alpha}
\newcommand{\be}{\beta}

\newcommand{\ga}{\gamma}

\newcommand{\rar}{\rightarrow}

\newcommand{\non}{\nonumber}

\allowdisplaybreaks[3]

\begin{document}

\title{Ultra-Compact accurate wave functions for He-like and Li-like iso-electronic sequences and variational calculus. III. Spin-quartet state of the Lithium sequence}

\author{D.J.~Nader}
\email{djuliannader@gmail.com}
\affiliation{Departamento de F\'isica, Universidad Veracruzana, C. P. 91090, Xalapa, Veracruz, M\'exico. }

\author{J.C.~del~Valle}
\email{delvalle@correo.nucleares.unam.mx}

\author{J.C.~Lopez Vieyra}
\email{vieyra@nucleares.unam.mx}

\author{A.V.~Turbiner}
\email{turbiner@nucleares.unam.mx, alexander.turbiner@stonybrook.edu}

\affiliation{Instituto de Ciencias Nucleares, Universidad Nacional Aut\'onoma de M\'exico,
A. Postal 70-543 C. P. 04510, Ciudad de M\'exico, M\'exico.}


\setstretch{1.4}

\begin{abstract}
As a continuation of Part I, dedicated to the ground state of He-like and Li-like isoelectronic sequences for nuclear charges $Z \leq 20$, and Part II, dedicated to two excited states of He-like sequence, two ultra-compact wave functions in the form of generalized Guevara-Harris-Turbiner functions are constructed for Li-like sequence. They describe accurately the domain of applicability of the Quantum Mechanics of Coulomb Charges (QMCC) for energies (2-3 significant digits (s.d.)) of the  spin-quartet state $1^40^+$   of Li-like ions (in static approximation with point-like, infinitely heavy nuclei). Variational parameters are fitted in $Z$ by 2nd degree polynomials. The most accurate ultra-compact function leads to the absolute accuracy $\sim 10^{-3}$\,a.u. for energy, and $\sim 10^{-4}$  for the normalized electron-nuclear cusp parameter for $Z \leq 20$.
Critical charge $Z=Z_B$, where the ultra-compact trial function for the $1^40^+$  state looses its square-integrability, is estimated, $Z_B(1^4\,0^+) \sim 1.26 - 1.30$. As a complement to Part I, square integrability for the compact functions constructed for the {\it ground, spin-doublet state} $1^2\,0^+$ of the Li-like sequence is discussed. The critical charge, for which these functions stop to be normalizable, is estimated as $Z_B( 1^2\,0^+)  = 1.62 - 1.65$. It implies that at $Z=2$ - the negative helium ion He$^-$ - both states $1^2\,0^+$ and $1^4\,0^+$ exist as states embedded to continuum.
\end{abstract}

\enlargethispage*{10pt}

\maketitle

\setstretch{1.5}

\section*{Introduction}

Ultra-compact approximate wave functions provide a valuable tool for different studies in
quantum mechanics. In the case of few-body Coulomb systems such compact functions allow us a simple physical interpretation of interactions (bonding mechanisms, in particular) between bodies in small atoms and molecules \cite{Harris:2005,David:2006,TGH:2009,H3+} and \cite{Part-1:2020,Part-2:2021}. Of course, this is in a great contrast with the highly complicated, lengthy variational trial functions composed by thousands terms, which are mainly aimed to get highly accurate (frequently excessively accurate, which never might be reached experimentally in nearest future) energies. In particular, it is an important case when the wave function includes non-linear parameters, those allow the physical meaning of charge screening, while their optimal values provide understanding of the physics picture of the Coulomb interactions in electronic media. Needless to say that compact trial functions are specially valuable when they are sufficiently accurate locally in the coordinate space being able to reproduce the energy as well as expectation values in a domain which is free of possible corrections of any type: finite mass, relativistic, QED etc. For the He-like and Li-like isoelectronic sequences the {\it correction-less} domain for the ground state energy was already localized in \cite{AOP:2019} by two of the present authors. It is of ``order" of 3-4 significant digits (s.d.). It is natural to assume this correction-less domain remains the same for energies of the excited states.

In Part I \cite{Part-1:2020} we introduced successfully the (ultra)-compact wave functions for He-like and Li-like iso-electronic sequences in their respective ground states with idea to get a description of the exact wave function inside the above-mentioned correction-less domain.  It has been already demonstrated that these functions can be successfully used for plasma-confined Helium atom \cite{Doma:2021}. In Part II \cite{Part-2:2021} similar (ultra)-compact wave functions for two low-lying 2${}^1\,S$ states and 1${}^3\,S$ of the He-like sequence were also  introduced.   These ultra-compact wave functions with a very few linear and non-linear variational parameters led not only to sufficiently accurate variational energies but also provide highly accurate expectation values and also the accurate description of cusp conditions. Furthermore, the non-linear parameters of those wave functions can be systematically described by the second degree polynomial in $Z$ for $Z \leq 20$ by making fit.
Square integrability of the ultra-compact functions for the He-like sequence for different $Z$ for three low-lying states was discussed additionally in Part II. As a result, critical charges $Z_B$ at which the compact wave functions stop to be normalizable (square-integrable) were found in each case. It was done for the ground state and two $S$ excited states for the He-like sequence. It was shown that $Z_B$ for the spin-singlet ground state and the first excited state coincide with high accuracy with each other and also with $Z_B$ for the ground state energy found in analysis of the Puiseux  expansion, see e.g. \cite{AOP:2019}. For spin-triplet state of the lowest energy $Z_B$ was found for the first time.

Needless to say that the low-lying excited states of Li atom and a few members of its isoelectronic sequence, were studied through the years using different approaches, mostly variational and Monte-Carlo ones. In particular, the accurate variational calculations of spin-quartet states of Li-like isoelectronic sequence of the type $(1s2s ns)$ $n=3,4,\ldots$ and $(1s2s np)$ $n=2,3,\ldots $ were performed in \cite{Holoien:1967,Barrois:1996,Barrois:1997}.
More recent calculations for the Lithium states $(1s\,2s\,ns)$ for $n=3-8$ and
$(1s\,2s\,np)$ for $n = 2-7$, for the cases of infinite and finite nuclear mass of ${}^7$Li, were done extensively in \cite{Yan:2003} by using the so-called double-basis set in Hylleraas coordinates of size $\sim (5000, 5000)$, where the benchmark energy value for the Li state $(1s2s3s)$  or ${1}^4\,S$ was found to be $E_{{1}^4\,S}^{\infty} = -5.212 748 247 225\,$a.u. (infinite nuclear mass) and $E_{{1}^4S}=-5.212 746 755 787\,$a.u. (finite nuclear mass).
Hence, the mass correction is very small: it changes the seventh s.d. being of the order of $\sim 10^{-6}$\,a.u.
As for another spin-quartet state $(1s2s2p)$ or ${1}^4\,P$  of the lowest energy the obtained value for the corresponding energy was $E_{{1}^4\,P}^{\infty} =-5.368 010 1539\,$a.u., and $E_{{1}^4\,P} =-5.368 025 6078\,$a.u. for infinite and finite nuclear mass, respectively. Mass correction for this state is $\sim 10^{-5}$\,a.u.
Comparison with the experimental results $E_{{1}^4\,S}^{\rm exp}= -5.2110\,$a.u. and $E_{{1}^4\,P}^{\rm exp}= -5.3660\,$a.u.\cite{NIST} indicates that corrections other than mass corrections are more important, since they can provide contributions $\sim 10^3$\,a.u. changing 3rd decimal digit (d.d.). Present authors are not aware about reliable systematic studies of relativistic and QED corrections for $Z > 3$, for discussion see \cite{AOP:2019}.

This paper is the third part in sequence: we follow the same strategy of building (ultra)-compact wave functions, as developed in Parts I,II\,, for zero total angular momentum - $S$ states - spin-quartet state  $1{}^4\,0^+$, or, in different nomenclature: $(1s\,2s\,3s)$ or ${1}^4\,S$ , of the lithium isoelectronic sequence in the domain of nuclear charges $Z \leq 20$.
Major emphasis in the construction of a compact wave function for Li-like systems is made to the insertion of electronic correlations in exponential form $\propto \exp{c_{ij}r_{ij}}$, where $r_{ij}$ is the relative distance between electrons $i,j (i\neq j) =1,2,3$ and $c_{ij}$ is the accompanying variational parameter and also natural modification $r_{ij} \rar {\hat r}_{ij}$ by replacing the distance by a rational function, see below. To the best of author's knowledge this form of correlation has been used in the past only for wave functions of the ground state of Li-like systems (see for instance \cite{TGH:2009}). This inclusion is particularly important in the domain of large distances approaching the asymptotics $r_j\to\infty$, since the exponential factors define the correct dominant behavior of the wave function at large distances. In particular, the exponential factors define the normalizability of the trial function.
Square integrability of the designed trial functions is studied as a function of the nuclear charge $Z$ and the critical charge at which these functions stop to be normalizable is determined. The same analysis of the square integrability and the determination of the associated critical charge is also carried out for the Li-like spin-doublet $1^20^+$ (ground) state ultra-compact functions constructed in Part I \cite{Part-1:2020}.

The structure of this paper is the following: in Section I, we briefly review the compact functions constructed for the spin-doublet (ground) state of the Li-like sequence, study their square integrability as a function of $Z$ and determine the critical charge $Z_B$. In Section II the Li-like sequence for the spin-quartet state $1{}^40^+$ is discussed, the exact solution for $Z \rar \infty$ is presented and (ultra)-compact functions are introduced which describe accurately the energy and the cusp conditions.
We then analyze the square integrability of state $1{}^40^+$ and the corresponding critical charge $Z_B$ is found. The results are summarized in Conclusions.

Atomic units are used throughout this paper.

\section{Li-like sequence: Spin-doublet state revisited.}

\subsection{Spin-doublet ground state: Generalities.}

Four compact trial functions for the spin $1/2$ ground state  $1^2 0^+$ of the lithium atom and its isoelectronic relatives ($3e;Z$) with infinite nuclear mass were considered and studied in details in Part I \cite{Part-1:2020}. The most accurate function among these ultra-compact functions taken as the variational trial function had provided an absolute accuracy for the energy $\sim 10^{-3}$ a.u. and from 2 to 3 s.d. for the electron-nuclear cusp parameter for $Z \leq 20$. Below we briefly review the three trial functions (a),(c),(d) used in Part I for the ground state and perform a detailed analysis of their normalizability for different $Z$. This analysis will allow us to estimate the critical charges $Z_B$ which mark the domain in $Z$ in where each of particular compact wave functions remains normalizable.

\subsection{Three compact trial functions for ground state.}

All compact variational functions which describe the ground state $1^2 0^+$ for Li-like sequence proposed in \cite{Part-1:2020} have the general form (see \cite{Part-1:2020} for details),
\begin{equation}
\label{GStrialfunct}
 \psi(\vec{r}_1,\vec{r}_2,\vec{r}_3; \chi) = {\cal A} \left[
   \,    \phi_1(\vec{r}_1,\vec{r}_2,\vec{r}_3) \chi_1 \ +\ C\,\phi_2(\vec{r}_1,\vec{r}_2,\vec{r}_3) \chi_2 \, \right]\, ,
\end{equation}
where $\chi_{1,2}$ stand for a 2-dimensional basis of the orthonormal three-electron spin 1/2 eigenfunctions. ${\cal A}$ is the three-electron antisymmetrizer
\begin{equation}
\label{Asym}
 {\cal A}\ =\ 1 - P_{12} - P_{13} - P_{23} + P_{231}  + P_{312}\, ,
\end{equation}
where $P_{ij}$ represents the permutation $(i \leftrightarrow j)$, and $P_{ijk}$ stands for the permutation of $(123)$ into $(ijk)$ \footnote{Note that the permutations $P_{231}$ and
$P_{312}$ correspond in {\it standard} notations to $P_{123}$ and $P_{132}$, respectively.}.
The parameter $C$ {\it measures} the relative contribution of each spin $1/2$ component in (\ref{GStrialfunct}). For the orbital part, the general Ansatz was given by the antisymmetrized three-electron exponentially correlated {\it seed}  functions $\phi_1,\phi_2$ (that depend only on relative distances)  of the Guevara-Harris-Turbiner form \cite{TGH:2009},
\begin{eqnarray}
\label{Phi-1}
\phi_1(\vec{r}_1,\vec{r}_2,\vec{r}_3; \al^{(1)}_i, \al^{(1)}_{ij}; a_3)  \ &=&\
 (1 - a_3 r_3+b_{12}r_{12})\
  e^{\scriptstyle -\al^{(1)}_1 Z r_1-\al^{(1)}_2 Z r_2-\al^{(1)}_3 Z r_3} \
   \non \\ & &\times \ e^{\scriptstyle \al^{(1)}_{12} {\hat r}_{12} + \al^{(1)}_{13} r_{13} + \al^{(1)}_{23} r_{23}} \ ,
\end{eqnarray}
and
\begin{eqnarray}
\label{Phi-2}
\phi_2(\vec{r}_1,\vec{r}_2,\vec{r}_3; \al^{(2)}_i, \al^{(2)}_{ij}; a_1)  \ &=&\
 (1 + a_1 r_{1}+b_{23}r_{23})\
 e^{\scriptstyle -\al^{(2)}_1 Z r_1-\al^{(2)}_2 Z r_2-\al^{(2)}_3 Z r_3} \
   \non \\ & &\times \ e^{\scriptstyle \al^{(2)}_{12} r_{12} + \al^{(2)}_{13} r_{13} + \al^{(2)}_{23} {\hat r}_{23}} \, ,
\end{eqnarray}
where $\al_i^{(p)}, \al_{ij}^{(p)}\,,\ j>i=1,2,3\ , p=1,2$\,, $a_1, a_3, b_{12}, b_{23}$, as well as  $C$ are considered as free variational
parameters.
Non-linear variational parameters $\al_i^{(p)}, \al_{ij}^{(p)}$ have a meaning of screening/antiscreening factors of the Coulomb charges in the nucleus-electron and electron-electron interactions, respectively.
%
Furthermore, the effects of charge screening,  which are assumed to be different at small and large distances, are taken into account by the following rational expressions inserted into the exponents of the Coulomb orbitals:
\begin{equation}
\label{substitution}
 \al_{i}r_{i}  \to   \al_{i}{\hat r}_{i}  \equiv   \al_{i}\,r_{i}\,\frac{1 + c_{i}  r_{i} }{1 + d_{i}  r_{i}} \ , \quad
 \al_{jk}r_{jk}  \to   \al_{jk}{\hat r}_{jk}  \equiv   \al_{jk}\,r_{jk}\,\frac{1 + c_{jk}\,  r_{jk}}{1 + d_{jk}\,r_{jk}}\ \ .
\end{equation}
Meaning of these expressions is interpolation the effective Coulomb charges between small and  large distances. For electron-nuclear attraction they are equal to $\al_{i}$, when $r_{i}$ is small, and to $\al_{i}\,\frac{c_{i}}{d_{i}}$, when $r_{i}$ is large. For electron-electron repulsion they are equal to $\al_{jk}$, when $r_{jk}$ is small, and to $\al_{jk}\, \frac{c_{jk}}{d_{jk}}$, when $r_{jk}$ is large.
In (\ref{substitution}) $c_i, d_i, c_{jk}$ and $d_{jk}$ are considered as variational parameters. Following physics arguments, described in Part I, the substitution (\ref{substitution})  was implemented in the terms
$\propto\exp(\al_{12}^{(1)}{\hat r}_{12})$ in $\phi_1$ (\ref{Phi-1}), and   $\propto\exp(\al_{23}^{(2)}{\hat r}_{23})$ in $\phi_2$ (\ref{Phi-2})
{\it only}. This conclusion was checked in variational calculations, where it was shown that the (anti)-screening in the form (\ref{substitution}) for other interactions does not effect
the first 3-4 s.d. in energy.
%

\subsection{Square integrability of trial functions for the ground state $1^2\,0^+$}
\label{squareintegrability}

It is evident that the normalizability of the entire wave function (\ref{GStrialfunct}) is guaranteed if seed functions are normalizable. For the seed wave functions $\phi_1$ (\ref{Phi-1}), $\phi_2$  (\ref{Phi-2})  to be  normalizable, they should decay exponentially  at large distances. In particular, this condition has to be valid whenever  the position of two electrons is kept fixed  while the position of the third electron tends to infinity. It can be explicitly seen in analysis of the exponential factors in (\ref{Phi-1}), (\ref{Phi-2}) that
\begin{equation}
\label{Adef}
\phi_{1,2} \big\vert_{r_i,r_j \ \rm fixed}  \to {\rm exp}\{- A_k^{(1,2)} r_k \}  \ \text{at} \  r_k\to\infty , \quad  i,j,k=1,2,3 \quad
(i\neq j\neq k)\ ,
\end{equation}
where the factors $A_k$ depend on $Z$, they are positive, $A_k^{(1,2)} > 0$, for large $Z > Z_B$. Also these factors can be ordered (see below) $$A_1^{(1,2)} > A_2^{(1,2)} > A_3^{(1,2)}\ .$$
These factors serve as a measure of the square integrability of $\phi_{1,2}$. Their inverses have a meaning of the average distance of the $k$-th electron to the nucleus, $\propto 1/A_k^{(1,2)}$. Our goal is to find the critical charge $Z_B$ when the smallest ($A_3$) vanishes.

In the limit $Z\to\infty$, where the electron-electron interactions can be neglected, the exact ground state seed wave function is the product of three anti-symmetrized Coulomb orbitals, and $A_k(Z)$ for $k$th electron grows linearly in $Z$:
\begin{equation}
\label{Ak0}
       A_k(Z) \underset{Z \to \infty}{=} \ \frac{1}{n}\, Z\ ,
\end{equation}
where $n$ takes values $n=1,2,3$ if the $k$th electron is in the $(1s)$, $(2s)$ or $(3s)$ state, respectively. In particular, if the 1st, 2nd and 3rd electrons are at the $(1s), (2s), (3s)$ states, respectively, then $A_1 > A_2 > A_3$. The quantities $A_k^{(1,2)}(Z)$ $(k=1,2,3)$ can be used to determine the value of the critical charge $Z=Z_B$ at which the trial function ($\phi_1$ and/or $\phi_2$) stops to be normalizable, $A(Z_B)=0$. This happens whenever one of the factor $A_1^{(1,2)}, A_2^{(1,2)}$ or $A_3^{(1,2)}$ vanishes first with $Z$ decreasing.
Concrete variational calculations show that the behavior of the quantities $A_{1,2,3}^{(1,2)}$ as  functions of the nuclear charge $Z$ exhibit essentially the linear behavior in domain where they are positive. In fact, any $A$ can be fitted with high accuracy by function of the form
\begin{equation}
\label{constraintsf}
A(Z)=b_{1/2}(Z-Z_b)^{1/2}+b_{1}(Z-Z_b)\,,
\end{equation}
where $b_{1/2}$, $b_{1}$ (the linear slope, c.f. (\ref{Ak0})), $Z_b>0$ (a critical charge for given $A$, for smallest $A$, $Z_b=Z_B$) are parameters. The square root term in (\ref{constraintsf}) is proposed by following the physics ``naturality": a singularity at the critical charge of $A(Z)$ is likely to be a square root branch point in $Z$-plane as it is in the ground state energy, see \cite{MPL:2016} and references therein. This form is motivated by the energy behavior observed in  He-like system (as well as many other systems of Coulomb charges) at the critical charge leading to appearance of the Puiseux expansion of the energy at $Z=Z_B$ (see \cite{MPL:2016} and references therein).  Furthermore, we found the  hierarchy $A_1(Z)>A_2(Z)>A_3(Z)$, which is certainly valid for $Z>Z_B$. It indicates the presence of a clusterization phenomenon which keeps electrons spatially separated, similar to what happens in He-like systems (see for discussion Part I).

\medskip

\subsection{Normalizability of the ground state functions. Results.}

\begin{figure}[htb]
\begin{center}
  \includegraphics[width=0.5\textwidth ,angle=-90]{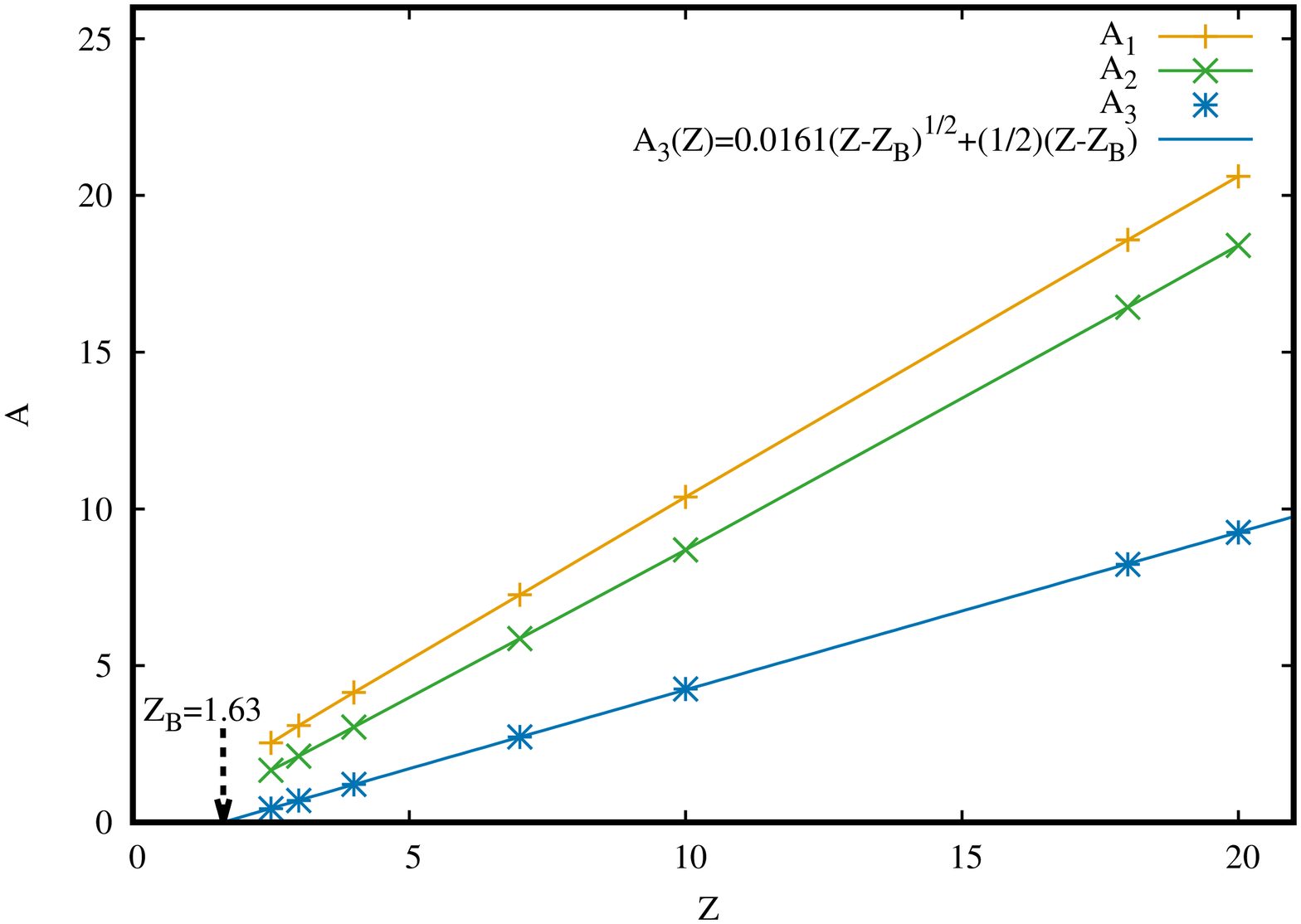}\\
  \includegraphics[width=0.5\textwidth ,angle=-90]{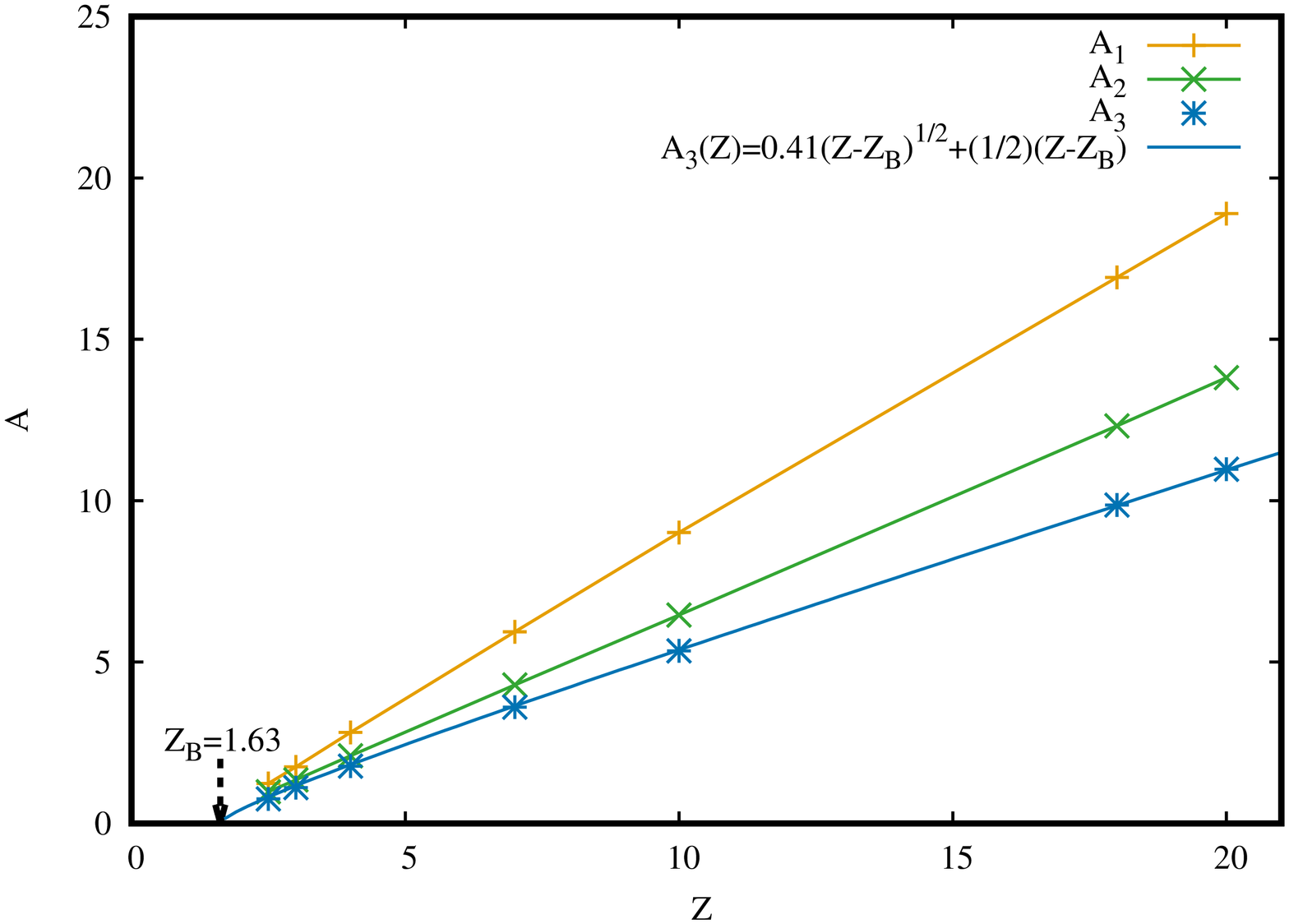}
  \caption{\label{GSAc}Ground state normalizability: $A_{1,2,3}$ for Li-like atoms {\it vs}  $Z$ for $\phi_1$ (top), and $\phi_2$ (bottom) of the Anzatz (c) (see Eq.(\ref{Aconstrc})). Normalizability is defined by the (smallest) $A_3$ leading to the critical charge $Z_B=1.63$. All $A_{1,2,3}$ can be fitted by a function of the form $A(Z)=a(Z-Z_B)^{1/2}+b(Z-Z_B)$ }
  \end{center}
\end{figure}

Among the trial functions considered in the study of the ground state $1^2\,0^+$ in Part I \cite{Part-1:2020} we choose the particular cases (a),(c),(d) of the general Ansatz (\ref{GStrialfunct}) for exploration in order to make analysis of their square integrability:

\begin{enumerate}

\item[{\bf (a)}] In the Ansatz (a) the orbital parts in (\ref{GStrialfunct}) are assumed equal, $\phi_1=\phi_2 \equiv \varphi$, they are the product of three correlated Hylleraas $(1s)$ type orbitals {\it i.e.} $\al_{i}^{(1)}=\al_{i}^{(2)} \equiv \al_i$,
    $\al_{ij}^{(1)}=\al_{ij}^{(2)}\equiv \al_{ij}$, $i=1,2,3$, $j<i$ and    $a_1=a_3=b_{12}=b_{23}=0$\,. Substitution (\ref{substitution}) is not implemented in the exponential correlation factors. The spin part of the function is separated out. In total, there are 7 variational parameters. It can be immediately seen that this wave function is normalizable, if all three expressions
\begin{eqnarray}
A_1^{(a)}&=& \al_1Z-\al_{12}-\al_{13}\ ,\non \\
A_2^{(a)}&=& \al_2Z-\al_{12}-\al_{23}\ , \label{Aconstra}\\
A_3^{(a)}&=& \al_3Z-\al_{13}-\al_{23}\ ,\non
\end{eqnarray}
are positive.  Here superindex $(a)$ stands for the seed function $\phi_1=\phi_2=\varphi$ of Ansatz (a). Since the variational parameters $\al_i,\al_{ij}$ have a smooth behavior as a function of the charge $Z$, the $A_{1,2,3}$ also have a smooth behavior {\it vs} $Z$.  The value of the critical charge is $Z=Z_B^{(a)}= 1.62$ at which the trial function (a) stops to be normalizable. This is determined by the value at which $A_3^{(a)}(Z)=0$, which is the smallest among $A_{1,2,3}^{(a)}$ for all $Z$ in the physical domain (see Fig. \ref{GSAa} below). In fact, $A_3^{(a)}(Z)$ is fitted like
\begin{equation}
\label{A3a}
    A_3^{(a)}(Z)\ =\ 0.0009\,(Z-Z_B^{(a)})^{1/2} + \frac{1}{2} (Z-Z_B^{(a)})\ ,
\end{equation}
with linear slope $b_1=1/2$ is fixed by the requirement that in the asymptotic limit $Z \rar \infty$ the third electron occupies the $(2s)$ orbital. Likewise, $A_1^{(a)}(Z), A_2^{(a)}(Z)$ are fitted by expressions of the type (\ref{A3a}) but with unit slope $b_1=1$. They play a secondary role since the normalizability is determined by $A^{(a)}_3(Z)$, see Fig.\ref{GSAa} below.

\item[{\bf (c)}]  In this Ansatz the orbital parts $\phi_1$ and $\phi_2$ in (\ref{GStrialfunct}) are different but they continue to be the product of two (modified) correlated Hylleraas $(1s)$ type orbitals and one correlated $(2s)$ orbital, thus, implying $b_{12}=b_{23}=0$. Note that substitution (\ref{substitution}) is not implemented in the exponential correlation factors. In total, the Anzatz (c) is characterized by 15 variational parameters including the parameter $C$.

The wave function for this case is normalizable while all the following six
expressions
\begin{eqnarray}
 A_1^{(c,i)}&=& \al_1^{(i)}Z-\al_{12}^{(i)}-\al_{13}^{(i)}\ ,\non \\
 A_2^{(c,i)}&=& \al_2^{(i)}Z-\al_{12}^{(i)}-\al^{(i)}_{23}\ ,
\label{Aconstrc} \\
 A_3^{(c,i)}&=& \al_3^{(i)}Z-\al_{13}^{(i)}-\al_{23}^{(i)}\ , \non
\end{eqnarray}
where $i=1,2$, remain positive. Here superindices $(c,1), (c,2)$ stand for the seed function $\phi_1,\phi_2$ of Ansatz (c), respectively.
Screening parameters $\al_i^{(1,2)},\al_{ij}^{(1,2)}$ behave smoothly {\it vs} $Z$ and so do the quantities $A_{1,2,3}^{(c_1,c_2)}$.
Expressions (\ref{Aconstrc}) are used to determine the value of the critical charge $Z=Z_B^{(c)}=1.63$ at which the trial function (c) stops to be normalizable. It turns out that the overall normalizability is defined by $A_3^{(c,1)}(Z)=0$ and $A_3^{(c,2)}(Z)=0$, which surprisingly(!) vanish almost simultaneously, for almost the same value of $Z$, corresponding to the loss of normalizability of $\phi_1$ (\ref{Phi-1}) and $\phi_2$ (\ref{Phi-2}), respectively. These expressions for $A$'s demonstrate a remarkable linear behavior with $Z$ almost everywhere with slope equals to $1/2$. In fact, they are accurately fitted by
\begin{align}
\label{Ac12}
 A_3^{(c,1)} &\ =\ 0.0161\,(Z-Z_B^{(c)})^{1/2}\ +\ \frac{1}{2} (Z-Z_B^{(c)})\ , \\
 A_3^{(c,2)} &\ =\ 0.41\,(Z-Z_B^{(c)})^{1/2}\ +\ \frac{1}{2} (Z-Z_B^{(c)})\ .
\end{align}
The expressions $A_1^{(c,1)}(Z)$ and $A_2^{(c,1)}(Z)$ are also approximately linear in $Z$ with slopes equal to $1$ in agreement with shell-model prediction: in $\phi_1$ electrons $1$ and $2$ are in $(1s)$ type orbitals (and thus $A^{(c,1;c,2)}$ is almost linear function with slope $1$) while the third electron is in the $(2s)$ orbital state demonstrating the linear behavior with slope $1/2$. On the other side, $A_1^{(c,2)}(Z)$ and  $A_2^{(c,2)}(Z)$ are also approximately linear with the slopes equal to $\sim 1$ and $\sim 3/4$, respectively, which indicates that in $\phi_2$ there is no definite arrangement of electrons distributed into the $(1s)$ and $(2s)$ orbitals (see Fig.\ref{GSAc}).

The value of the critical charge for this function $Z_B^{(c)}=1.63$ is slightly larger than one found in the case (a). This is expected since the wave function (c) contains more parameters and, following our philosophy, represents more accurate approximation to the exact wave function.

\bigskip

\item[{\bf (d)}]  In this Ansatz,  the orbital parts in (\ref{GStrialfunct}) are assumed to be different, both $\phi_1$ and $\phi_2$ are made as the product of (modified) Guevara-Harris-Turbiner  functions, {\it i.e.} all variational parameters in (\ref{GStrialfunct}) are free, and additionally the interpolation (\ref{substitution}) is inserted into the Coulomb exponential correlation terms, $\sim \exp (\alpha_{12}^{(1)} {\hat r}_{12})$ and $\sim \exp (\alpha_{23}^{(2)} \hat{r}_{23})$, to assure more accurate screening. In total, we have 21 variational  parameters including the parameter $C$, see Eq.(1).

    Expressions (\ref{Adef}) for the exponential factors indicate that the wave function for this case is normalizable when all
\begin{eqnarray}
  A_1^{(d_1)}& = & \al_1^{(1)}Z-\al_{12}^{(1)}\frac{c_{12}^{(1)}}{d_{12}^{(1)}}-\al_{13}^{(1)}\,,\non \\
  A_2^{(d_1)}& = & \al_2^{(1)}Z-\al_{12}^{(1)}\frac{c_{12}^{(1)}}{d_{12}^{(1)}}-\al_{23}^{(1)}\,,\non \\
  A_3^{(d_1)}&=& \al_3^{(1)}Z-\al_{13}^{(1)}-\al_{23}^{(1)}\,,\label{Aconstrd} \\
  A_1^{(d_2)}&=& \al_1^{(2)}Z-\al_{12}^{(2)}-\al_{13}^{(2)}\ ,\non \\
  A_2^{(d_2)}&=& \al_2^{(2)}Z-\al_{12}^{(2)}-\al_{23}^{(2)}\frac{c_{23}^{(2)}}{d_{23}^{(2)}}
  \ ,\non \\
  A_3^{(d_2)}&=& \al_3^{(2)}Z-\al_{13}^{(2)}-\al_{23}^{(2)}\frac{c_{23}^{(2)}}{d_{23}^{(2)}}\ , \non
\end{eqnarray}
cf.(\ref{Aconstra}), (\ref{Aconstrc}),
are positive. Here, the superindices $(d_1), (d_2)$ stand for the exponential factors orbitals of seed function $\phi_1,\phi_2$ of Ansatz (d) respectively.
Quantities (\ref{Aconstrd}) are used to determine the value of the critical charge $Z=Z_B^{(d)}=1.65$ at which the corresponding trial function stops to be normalizable. This value represents a better approximation to the exact critical value.  It turns out that the overall normalizability is defined by $A_3^{(d_1)}(Z)=0$ or $A_3^{(d_2)}(Z)=0$ since they vanish simultaneously, and  are fitted by
\begin{align}
  A_3^{(d_1)} &\ =\ 0.012\,(Z-Z_B^{(d)})^{1/2}\ +\ \frac{1}{2} (Z-Z_B^{(d)})\ , \\
  A_3^{(d_2)} &\ =\ 0.53\, (Z-Z_B^{(d)})^{1/2}\ +\ \frac{1}{2} (Z-Z_B^{(d)})\ .
\end{align}
The relations $A_1^{(d_1)}(Z)$ and  $A_2^{(d_1)}(Z)$ are also approximately linear with slope equal to  $1$ in agreement with the shell model picture. On the other side, $A_1^{(d_2)}(Z)$ and  $A_2^{(d_2)}(Z)$ are also approximately linear but the slopes 1 and 3/4, respectively, see Fig. \ref{GSAd}.
\end{enumerate}

In general, the optimal variational parameters can be fitted by simple quadratic functions of the nuclear charge $Z$, with the only exception of parameter $C$ which is fitted by a quadratic function on $Z^{-2}$, see Part I for comparison.
In particular, the quadratic in $Z$ fits corresponding to the parameters of Anzatz (d) are
{
\begin{eqnarray}
 C & = & 0.0206 - 0.7041Z^{-2} + 35.4617Z^{-4} \ , \non \\
 a_3 & = & 0.3749 + 0.5262 Z - 0.0014 Z^2 \ ,\non \\
 b_{12} & = & 0.1275-0.0338Z+0.0007Z^2 \ , \non \\
 \al_1^{(1)} Z & = & 0.0953 + 1.0775 Z -0.0013 Z^2 \ ,\non \\
 \al_2^{(1)} Z & = & -0.4456 + 0.9293 Z + 0.0012 Z^2 \ ,\non \\
  \al_3^{(1)} Z & = & -0.8266 + 0.5193 Z - 0.0004 Z^2 \ ,\non \\
  \al_{12}^{(1)} & = & 0.3340 + 0.0005 Z  -0.000006 Z^2 \ ,
\label{fits_GS_d} \\
  c_{12}^{(1)} &=&  0.0659 -0.0720 Z  +0.0009 Z^2 \ , \non \\
  d_{12}^{(1)} &=& 0.0005 + 0.0032 Z + 0.0053 Z^2 \ ,\non \\
  \al_{13}^{(1)} &=& 0.0098 -0.0017 Z  +0.000006 Z^2 \ ,\non \\
  \al_{23}^{(1)} &=&  -0.0105+0.0152 Z  -0.0002 Z^2 \ ,\non \\
  a_1  &=&  0.5358 - 0.2054 Z +0.0025 Z^2  \ ,\non \\
  b_{23}  &=& 0.1062 -0.0293 Z  -0.00001 Z^2 \ ,\non \\
  \al_{1}^{(2)} Z &=& -0.3014 + 1.0156Z  -0.0028 Z^2 \ ,\non \\
  \al_{2}^{(2)} Z &=&  -0.2095 + 0.8899Z  - 0.0005Z^2 \ ,\non \\
  \al_{3}^{(2)} Z &=&  1.1304 +  0.2485Z  + 0.0049 Z^2 \ ,\non \\
  \al_{12}^{(2)} &=& -0.0300+  0.2100 Z   -0.0036 Z^2 \ ,\non \\
  \al_{13}^{(2)} &=& 1.2212 -0.3388 Z  + 0.0053 Z^2\ , \non \\
  \al_{23}^{(2)} &=& 0.1754 +0.0461 Z  -0.0004 Z^2 \ ,\non \\
  c_{23}^{(2)} &=& 0.0560 - 0.0018 Z + 0.00006 Z^2 \ , \non \\
  d_{23}^{(2)} &=& - 0.0010 -0.0108 Z + 0.0087 Z^2 \ , \non
\end{eqnarray}}
cf. Table 6 in Part I \cite{Part-1:2020}.
By taking the concrete values of the fitted parameters one can get, in general, the energies with accuracy $\sim 3$ s.d. \footnote{Note that the quadratic fits presented in (\ref{fits_GS_d}) are slightly different (in some cases) then those presented in the Part I, but the variational energy remains unchanged up to 4 s.d. It allows to have (slightly) smoother behavior of the functions $A_k, k=1,2,3$.}

It is worth noting that stemming from the analysis of the normalizability of the wave functions (a), (c) and (d) {\it vs  $Z$} the ground state for negative helium ion ${\rm He}^-$  ($Z=2$) remains still normalizable since $A_3(Z=2)>0$ even though the energy ($E_{{\rm He}^-}=-2.8992\,$ Hartrees) is larger than that one corresponding to system ${\rm He} + e$. Thus, all wave functions considered in Part I for the Li-like systems describe ${\rm He}^-$ as a bound state embedded in the continuum. The interesting question what would happen if more accurate function is taken as the trial function, will this level become the bound state with energy below threshold - it is not clear to the present authors, it might be a subject of separate study.

\begin{figure}[htb]
\begin{center}
  \includegraphics[width=0.5\textwidth ,angle=-90]{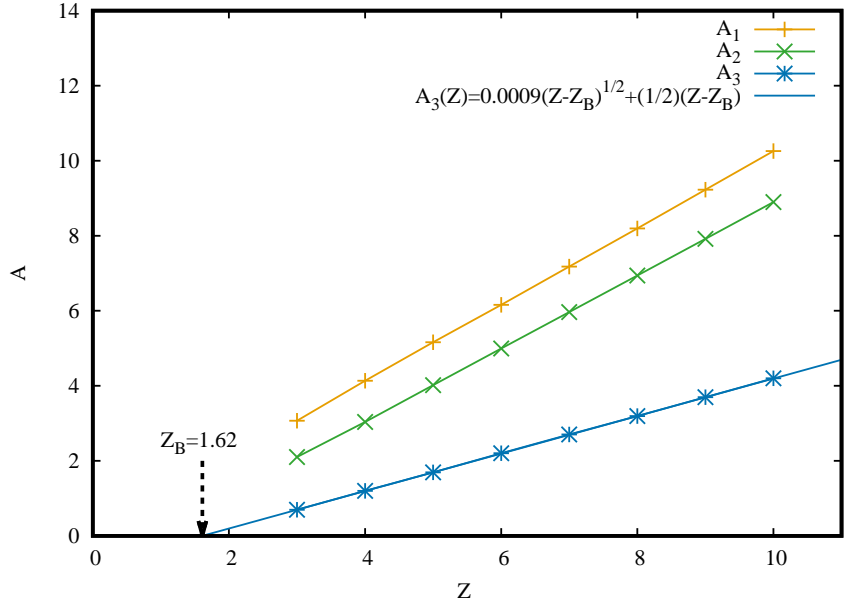}
  \caption{\label{GSAa}Ground state normalizability: $A_{1,2,3}^{(a)}$ for Li-like atoms as function of the nuclear charge $Z$ for the trial wave function (a) (see eq. (\ref{Aconstra})).
  All $A_{1,2,3}^{(a)}$ fitted by functions of the form (\ref{constraintsf}).
  The overall normalizability of (a) defined by the (smallest) $A_3^{(a)}$ leading to the critical charge \hbox{$Z_B^{(a)}=1.62$}. As for Anzatzen (c) and (d) it leads to critical charges \hbox{$Z_B^{(c)}=1.63$} and \hbox{$Z_B^{(d)}=1.65$}, see below Figs. \ref{GSAc} and \ref{GSAd}, respectively.
  }
  \end{center}
\end{figure}

\begin{figure}[htb]
\begin{center}
  \includegraphics[width=0.6\textwidth ,angle=-90]{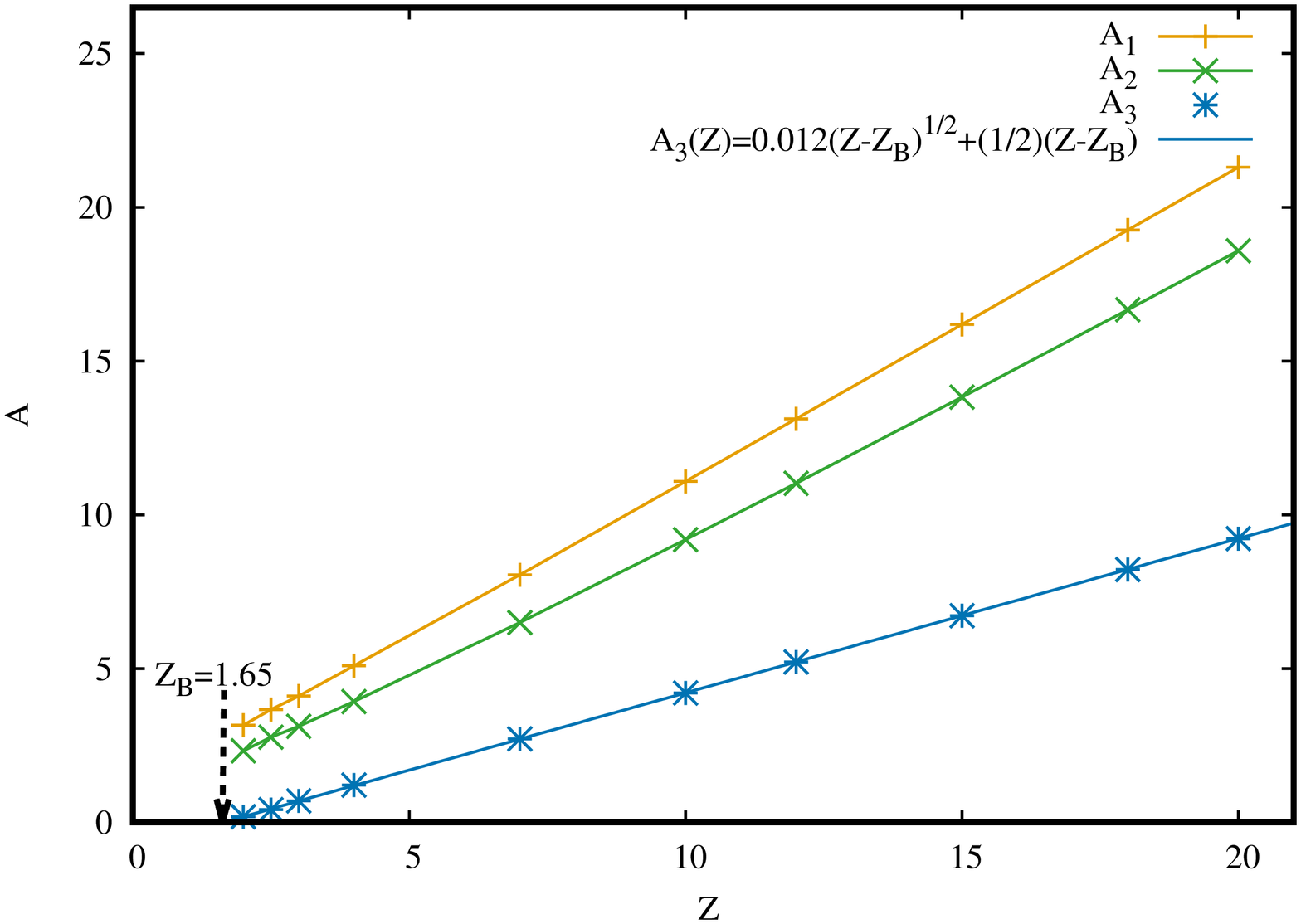}\\
   \includegraphics[width=0.6\textwidth ,angle=-90]{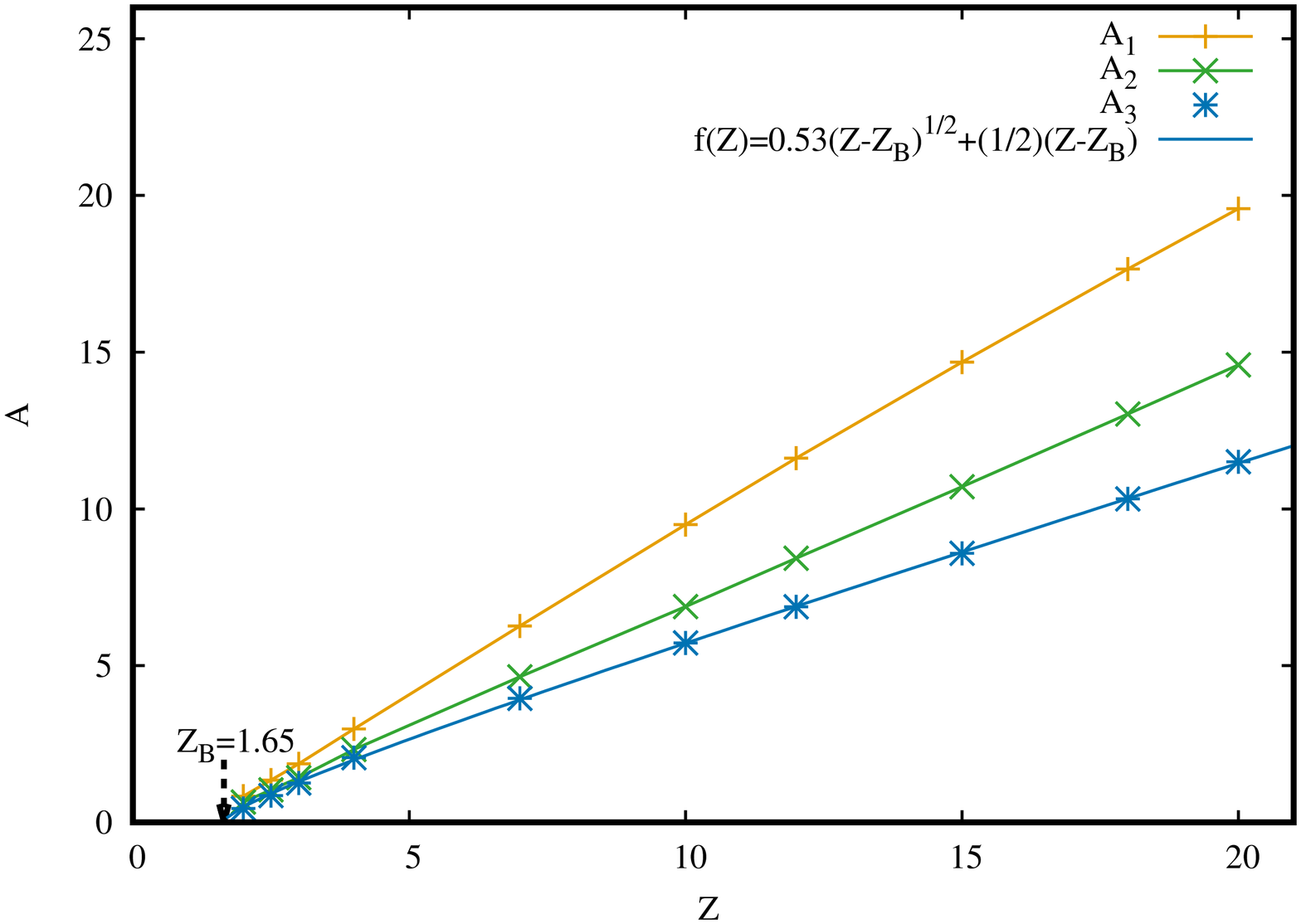}
  \caption{\label{GSAd}Ground state normalizability: $A_{1,2,3}$ for Li-like atoms {\it vs} $Z$  for the orbital functions $\phi_1$ (top), and $\phi_2$ (bottom) of the Anzatz (d) (see Eq.(\ref{Aconstrd})). Normalizability defined by the (smallest) $A_3$ leading to the critical charge $Z_B=1.65$. All $A_{1,2,3}$ can be fitted by a function of the form $A(Z)=a(Z-Z_B)^{1/2}+b(Z-Z_B)$.
  }
\end{center}
\end{figure}

\section{ Li-like sequence: spin-quartet state}

\subsection{Generalities. Exact solution for the spin-quartet state.}

For the Lithium-like system $(3e; Z)$ with infinite nuclear mass, we look for a compact function which would describe the lowest energy $S$ state $1{}^40^+$ ($1^4 S$) of total spin
$3/2$ in the form
\begin{equation}
\label{EStrialfunct}
 \psi(\vec{r}_1,\vec{r}_2,\vec{r}_3; \chi) = {\cal A} \left[
   \,\varphi(\vec{r}_1,\vec{r}_2,\vec{r}_3)\, \chi \, \right]\ ,
\end{equation}
cf.(\ref{GStrialfunct}), where $\chi$ represents a spin 3/2 eigenfunction made from three spin-1/2 electronic states with zero total angular momentum.  It implies that all electrons are in the same spin state. The operator ${\cal A}$ is the three-electron antisymmetrizer (\ref{Asym}).  In a similar way as for the ground state (\ref{GStrialfunct}), the function $\varphi$ in (\ref{EStrialfunct}) is called the {\it seed} function. As a result of antisymmetrization the function $\psi$ contains six terms.


At $Z \to \infty$, where the electron-electron interaction disappears, the problem is reduced to that of three Hydrogen atoms in (different) $s$-orbitals. In this limit, there exists the exact solution of the Schr\"odinger equation in the form of anti-symmetrized product of three Coulomb orbitals $(1s\,2s\,3s)$ with the seed function
\begin{equation}
\label{Li-infty-psi-}
  \varphi_0\ =\ (1 - a Z r_2) (1 + b Z r_3 +  c Z^2 r_3^2)\,
  e^{-\al_1 Z r_1 - \al_2 Z r_2 - \al_3 Z r_3}\ \sim (1s_1\, 2s_2\, 3s_3) \ ,
\end{equation}
with $\al_1=1, a=\al_2=1/2,\ b = -2/3, c =  2/27, \al_3=1/3$. It corresponds to the energy
\[
     E_0=-\frac{49}{72}\,Z^2 \ .
\]
The function (\ref{Li-infty-psi-}) can be used as the trial function with $\al_{1,2,3},a,b,c$ taken as variational parameters. This function represents the starting point to construct the compact trial functions, it will be presented below.

\subsection{Compact Trial Functions}

Let us present three (ultra)-compact trial functions for the spin-quartet state $1\,{}^4 0^+$ (${1}^4S$) in addition to (\ref{Li-infty-psi-}).

\begin{itemize}

\item[(i)] The first (ultra)-compact variational function is constructed by making a generalization of the function (\ref{Li-infty-psi-}) by adding the electronic correlations in the form of exponentially correlated {\it Hylleraas-type} functions
\begin{equation}
\label{Li-psi-1}
  \varphi_1\ =\ (1 - a^{(1)} Z r_2) (1 + b^{(1)} Z r_3 +  c^{(1)} Z^2 r_3^2)\,
  e^{-\al_1^{(1)} Z r_1 - \al_2^{(1)} Z r_2 - \al_3^{(1)} Z r_3 + \al_{12}^{(1)} r_{12} + \al_{13}^{(1)} r_{13} + \al_{23}^{(1)} r_{23}}\ ,
  \end{equation}
where $\al_{1,2,3}^{(1)}$ and $\al_{12}^{(1)}, \al_{13}^{(1)}, \al_{23}^{(1)}$ are variational parameters having the meaning of screening factors of the charges in the electron-nucleus interactions and electron-electron interactions, respectively.
Note that the prefactor in (\ref{Li-psi-1}) is chosen in the same functional form as in the exact solution (\ref{Li-infty-psi-}).
The parameters $a^{(1)},b^{(1)},c^{(1)}$ which appear in the exact  $(2s), (3s)$ prefactors in (\ref{Li-infty-psi-}) are also treated as variational parameters.
In total, this function is characterized by 9 free parameters; it will be denoted as the Anzatz (a).
The results of the variational energy using this function for charges in the range $2 < Z \leq 20$ are shown in Table  \ref{resultsli} (marked with a label ${}^{(a)}$), together with the value of the normalized nucleus-electron cusp parameter.  Compared to various accurate results found in the literature (see for instance \cite{King}, \cite{QMC:2014}, \cite{NIST}), this trial function reproduces 2-3 s.d. in the energy for all studied values of $Z \leq 20$. Note that the presented variational energies are systematically lower than those obtained in the Hartree-Fock formalism. It is remarkable that the variational trial function (\ref{Li-psi-1}) leads to normalized nucleus-electron cusp values which are close to the exact value equal to 1. For details on the numerical procedure the reader is referred to Section 3.1 of Part I \cite{Part-1:2020}.

\item[(ii)] As for the second function the variational function (\ref{Li-psi-1}) can be
           further generalized to the form
\begin{align}
\label{Li-psi-2}
  \varphi_2\ &=\ (1 - a_1^{(2)} Z r_2 + b_{13}^{(2)} r_{13}) (1 - a_2^{(2)} Z r_3 + a_3^{(2)} Z r^2_3 + b_{12}^{(2)} r_{12})\,  \non \\[5pt]
  & \qquad\qquad \times \ e^{-\al_1^{(2)} Z {\hat r}_1 - \al_2^{(2)} Z {\hat r}_2 - \al_3^{(2)} Z {\hat r}_3 + \al_{12}^{(2)} {\hat r}_{12} + \al_{13}^{(2)} {\hat r}_{13} + \al_{23}^{(2)} {\hat r}_{23}}\ ,
\end{align}
where into prefactors for the second electron in $(2s)$ state (linear in $r_2$) and the third electron in $(3s)$ state (quadratic in $r_3$) are introduced terms $\sim b_{13}^{(2)} r_{13}$ and $\sim b_{12}^{(2)} r_{12}$, respectively, where  $b_{13}^{(2)},b_{12}^{(2)}$ are variational parameters. Additionally, in order to have a more adequate description of the Coulomb interactions between each pair of electrons at small and large distances, an interpolation of the effective Coulomb interactions (\ref{substitution}) was introduced into the correlation factors, $\exp(\al_{jk}^{(2)}{\hat r}_{jk})$,  $jk= 12, 13, 23$.  The same substitution into the Coulomb $s$-orbitals  $ \exp(- \al_i^{(2)} Z {\hat r}_i)$, $i=1,2,3$   was done to interpolate the Coulomb interaction of the nucleus with electrons at small and large distances.  Concrete calculations indicate that a similar substitution into the modified $(1s)$ Coulomb orbital for the first electron is, in fact, unnecessary for all $Z \leq 20$: it leads to a change in variational energy beyond of the fourth s.d. Similar situation occurs for $(3s)$ Coulomb orbital for the third electron, see below. Details on the numerical calculations with this trial wave function, are similar ones of the ground state described in Section 3.1 in Part I \cite{Part-1:2020}.

\item[(iii)] Concrete variational calculations using (\ref{Li-psi-2}) show that for all studied $Z$ the values of parameters $b_{13}^{(2)}, b_{12}^{(2)}$, $c_3^{(2)},d_{3}^{(2)},c_{13}^{(2)}$ and $d_{13}^{(2)}$ are small in comparison with other parameters: without loosing the accuracy in the first 4 s.d. in energy they can be placed equal to zero. Therefore, one can reduce effectively the number of parameters in (\ref{Li-psi-2}) without an essential deterioration in the energy inside the correction-less domain.  The correlation terms $\propto r_{ij}$ in the prefactors do not play an important role similar as it was for He-like systems. Eventually, this consideration leads to the final, third Ansatz for seed function in the form
\begin{align}
  \varphi_3\ &\ =\   (1 - a^{(3)} Z r_2) (1 + b^{(3)} Z r_3 +  c^{(3)} Z^2 r_3^2)\,
\label{Li-psi-3} \\
  & \qquad\qquad \times e^{ -\al_1^{(3)} Z r_1 - \al_2^{(3)} Z {\hat r}_2 - \al_3^{(3)} Z r_3 + \al_{12}^{(3)} {\hat r}_{12} + \al_{13}^{(3)}  r_{13} + \al_{23}^{(3)} {\hat r}_{23}}\ .
\non
\end{align}
It will be used throughout and denoted as the Anzatz (b). This function contains 15 variational parameters: linear parameters $a^{(3)},b^{(3)},c^{(3)}$ in the prefactor, non-linear parameters $\al_1^{(3)},\al_2^{(3)}, \al_3^{(3)}, \al_{12}^{(3)}, \al_{13}^{(3)}, \al_{23}^{(3)}$ in the exponential terms, plus the non-linear parameters for the interpolations (\ref{substitution}):
$ c_2^{(3)},d_2^{(3)},  c_{12}^{(3)},d_{12}^{(3)},  c_{23}^{(3)},d_{23}^{(3)}$. This
trial function is the most accurate among the trial functions of the type (\ref{Li-psi-2}), it hints that the structure of the prefactors is the same as that occurring in the exact wave function $(1s\,2s\,3s)$ in the limit $Z \to \infty$. In particular, this means that the addition of correlation terms $\propto r_{jk}$ in prefactor does not play a significant role in the design of trial functions for the state $1{}^4\,0^+$ ($1{}^4S$) unlike the spin-doublet ground state, see Part I \cite{Part-1:2020} and also \cite{TGH:2009}.
It seems that this systematics can hold for other excited states when building compact trial functions.

The variational energies obtained the Anzatz (b) (\ref{Li-psi-3}) for $2 \leq Z \leq 20$ are shown in Table \ref{resultsli} (marked with the label ${}^{(b)}$), together with the value of the normalized nucleus-electron cusp parameter.  Compared to more accurate variational results by King \cite{King} and those based on Monte-Carlo calculations \cite{QMC:2014}, as well as experimental ones \cite{NIST}, the trial function (b) reproduces 2-3 s.d. correctly in energies. It is remarkable that the variational results yield normalized nucleus-electron cusp parameters close to the exact value 1.  Note that the improvement in the energy coming from the trial function (b) with respect to the values obtained with the trial function (a) occurs in the 4-5 s.d.
\end{itemize}

\subsection{Square Integrability of compact wave functions for the spin-quartet state $1\,{}^4\,0^+$}

The analysis of the square integrability of the trial functions, constructed in previous Section for the Li-like $1\,{}^4\,0^+$ spin-quartet state, follows the same lines as the study of the  square integrability of the $1^2\,0^+$ (ground) state trial functions discussed in Section I.C, see above.  In particular, the relations (\ref{Adef}) for the exponential factors indicate that the  square integrability of the trial wave function (a) with seed function $\varphi_1$ (\ref{Li-psi-1}) is guaranteed as long as the following quantities
\begin{eqnarray}
A_1^{(\varphi_1)}&=& \al_1^{(1)}Z-\al_{12}^{(1)}-\al_{13}^{(1)}\ ,\non \\
A_2^{(\varphi_1)}&=& \al_2^{(1)}Z-\al_{12}^{(1)}-\al_{23}^{(1)}\ ,
\label{constraintsa} \\
A_3^{(\varphi_1)}&=& \al_3^{(1)}Z-\al_{13}^{(1)}-\al_{23}^{(1)}\ , \non
\end{eqnarray}
remain positive. Likewise, the  square integrability of the trial wave function (b) with seed function $\varphi_3$  (\ref{Li-psi-3}) included is guaranteed as long as the following quantities
\begin{eqnarray}
A_1^{(\varphi_3)}&=& \al_1^{(3)}Z-\al_{12}^{(3)}\frac{c_{12}^{(3)}}{d_{12}^{(3)}}-\al_{13}^{(3)}\,,\non \\
A_2^{(\varphi_3)}&=& \al_{2}^{(3)}Z\frac{c_2^{(3)}}{d_2^{(3)}}-\al_{12}^{(3)}\frac{c_{12}^{(3)}}{d_{12}^{(3)}}-\al_{23}^{(3)}\frac{c_{23}^{(3)}}{d_{23}^{(3)}}\,,\label{constraintsb} \\
A_3^{(\varphi_3)}&=& \al_3^{(3)}Z-\al_{13}^{(3)}-\al_{23}^{(3)}\frac{c_{23}^{(3)}}{d_{23}^{(3)}}\,, \non
\end{eqnarray}
remain positive.

In Fig. \ref{constraintsES} we present  the behavior  of the quantities $A_i^{(\varphi_1)}$ and $A_i^{(\varphi_3)}$, $i=1,2,3$ as a function of $Z$, corresponding to the optimal parameters of the trial functions generated from the seed functions $\varphi_1$ and $\varphi_3$ respectively.  All of these quantities present a linear behavior as functions of $Z$ and permit to get an estimate for the critical charges $Z=Z_B$ at which the trial functions stop to be normalizable (see below).

The optimal variational parameters  in (\ref{Li-psi-2}) and (\ref{Li-psi-3}),  which lead to the minimal variational energy, have a smooth behavior as functions of the nuclear charge
$Z$. They can be fitted easily by second degree polynomials in $Z$. The fits corresponding to the most accurate Ansatz (b) with seed function $\varphi_3$ are the following
\begin{eqnarray}
  a &\ =\ & 0.2354+0.5693 Z - 0.0023 Z^2\ ,\non \\
  b &\ =\ & 2.1011-0.1118 Z+0.0117Z^2\ ,\non \\
  c &\ =\ & 1.9858+0.2292 Z+0.0030Z^2\ ,\non \\
  \al_1 Z  &\ =\ & 0.0076 + 0.9994 Z - 0.000001 Z^2\ ,\non \\
  \al_2 Z  &\ =\ & -0.2591 + 0.4839 Z + 0.0004 Z^2\ ,\non \\
  c_2  &\ =\ & -0.0370 + 0.0146 Z - 0.0003 Z^2\ ,\non \\
  d_2  &\ =\ & -0.0337 + 0.0133 Z - 0.0003 Z^2\ ,\non \\
  \al_3 Z  &\ =\ & -0.4416 + 0.3334 Z - 0.0003 Z^2
\label{fits_ES_beta} ,\\
 \al_{12} &\ =\ & 0.0369 + 0.0225 Z  -0.0005 Z^2\ ,\non \\
  c_{12} &\ =\ &  0.0962 - 0.0466 Z  -0.0002 Z^2\ ,\non \\
  d_{12} &\ =\ & -0.1642 + 0.0776 Z + 0.0004 Z^2\ ,\non \\
  \al_{13} &\ =\ & 0.0130 -0.0001 Z  -0.00008 Z^2\ ,\non \\
  \al_{23} &\ =\ &  0.2196 -0.0054 Z  +0.0007 Z^2\ , \non \\
   c_{23} &\ =\ &  0.0072 -0.0020 Z  +0.00005 Z^2\ ,\non \\
   d_{23} &\ =\ & 0.1002 -0.0130 Z + 0.0034 Z^2 \ . \non
 \end{eqnarray}
Optimal variational parameters, in general, lead to energies having an accuracy of 3 s.d. compared with most accurate results known in literature. The use of the fitted parameters does not deteriorate the variational energy: the difference occurs beyond the 4th decimal digit.

\begin{table}[]
  \caption{\label{resultsli} Energy (in Hartrees) of Li-like ions for the state $1\,{}^40^+$ for nuclear charges $3 \leq Z \leq 20$ obtained with (\ref{EStrialfunct}):
  $^{(a)}$ with seed function $\varphi_1$ (\ref{Li-psi-1}), $^{(b)}$ with seed function $\varphi_3$  (\ref{Li-psi-3}). The column marked as $E_{M}$ stands for energies  from Majorana formula (\ref{Li-majorana}). The most accurate result for Li-atom \cite{King} marked by $\dagger$, $HCI$ (Hylleraas CI) from \cite{Barrois:1996}, $HF$ (Hartree-Fock), $FCI$ (Full CI) and  DMC  (Diffusion Monte Carlo) taken from \cite{QMC:2014} and $E^{exp}$ experimental energies from \cite{NIST}.
  Normalized cusp parameters $\mathcal{C}_{Ne}$ are included.}
\label{resultsl}
\begin{center}
         {\setlength{\tabcolsep}{0.5cm}
                \renewcommand{\arraystretch}{0.7} 
                 \resizebox{0.9\textwidth}{!}{%
\begin{tabular}{ccccclc}
\hline
     $\quad Z$ & $E^{(a)}$(a.u.) & $\mathcal{C}^{(a)}_{Ne}$ &$E^{(b)}$(a.u.)&
     $\mathcal{C}_{Ne}^{(b)}$ & $E_{ref}$(a.u.)& $E_{M}$(a.u.)  \\
\hline
              \rule{0pt}{4ex}2  &   & & {\it -2.1656} & {\it 1.005} &  &
                         {\it  -2.1641} \\
              \rule{0pt}{4ex}3  & -5.2088 & 1.005 & -5.2095 & 1.003 & -5.212748$^{\dagger}$ & -5.2080\\
              \rule{0pt}{4ex}   & & & & &-5.2036$^{HF}$    \\
              \rule{0pt}{4ex}   & & & & &-5.21098$^{DMC }$ \\
              \rule{0pt}{4ex}   & & & & &-5.21275$^{HCI}$  \\
              \rule{0pt}{4ex}   & & & & &-5.2113$^{FCI}$   \\
              \rule{0pt}{4ex}   & & & & &-5.2110$^{exp}$   \\
              \rule{0pt}{4ex}4 &  -9.6121 & 1.004 & -9.6129 & 1.003 & -9.6198$^{HCI}$
              & -9.6131 \\
                          \rule{0pt}{4ex}   & & & & &-9.6207$^{exp}$     \\
                          \rule{0pt}{4ex}5 & -15.3777 & 1.003 & -15.3786 & 1.002 & -15.3710$^{HF}$ &-15.3792 \\
                          \rule{0pt}{4ex}   & & & & &-15.38898$^{DMC }$     \\
                          \rule{0pt}{4ex}   & & & & &-15.3895$^{HCI}$     \\
                          \rule{0pt}{4ex}   & & & & &-15.3895$^{FCI}$    \\
                          \rule{0pt}{4ex}   & & & & &-15.3934$^{exp}$    \\
            \rule{0pt}{4ex}9 & -52.0534 & 1.002         & -52.0549 & 1.001 & -52.0459$^{HF}$ &-52.0549\\
            \rule{0pt}{4ex}   & & & & &-52.08257$^{DMC }$  \\
            \rule{0pt}{4ex}   & & & & &-52.0827$^{HCI}$    \\
            \rule{0pt}{4ex}   & & & & &-52.0832$^{FCI}$    \\
            \rule{0pt}{4ex}10 & -64.6254& 1.001 &-64.6267 & 1.001 & -64.6591$^{HCI}$
            & -64.6266 \\
            \rule{0pt}{4ex}15 & -147.9002 & 1.001 & -147.9028 & 1.001 & - & -147.9018\\
            \rule{0pt}{4ex}18 & -214.1983 & 1.001 & -214.2005 & 1.001 & -214.1946$^{HF}$ & -214.2003\\
            \rule{0pt}{4ex}   & & & & & -214.27009 $^{DMC}$    \\
            \rule{0pt}{4ex}20 & -265.1994 & 1.000 & -265.2047 & 1.001 & - & -265.2048 \\[5pt]
\hline
\end{tabular}}
}
\end{center}
\end{table}

\begin{table}
\caption{\label{table2} Spin-quartet state $1^40^+$ (${1}^4S$) of the Li-like sequence,
         expectation values of:\\ (i)  the potential $\langle V \rangle$  in a.u.,
         (ii) $\langle r_{eN}^2 \rangle$  in (a.u.)$^2$, and (iii) $\langle r_{12} \rangle$ in a.u. {\it vs} $Z$ obtained with the trial function (\ref{EStrialfunct}):
         $^{(a)}$ with seed function $\varphi_1$  (\ref{Li-psi-1}),
         $^{(b)}$  with seed function $\varphi_3$  (\ref{Li-psi-3})
         compared with $\langle V \rangle$ from \cite{Barrois:1996}.
         Columns 7,8 {\color{red} \bf (??)}: the expectation value $\langle r_{eN}^2 \rangle$ for the ground state $1^2 0^+$ from Ansatz (d) of (\ref{GStrialfunct}) -- see the text, compared with \cite{Yan:1995} ${}^\dagger$ and \cite{Frolov:2014} ${}^\ddagger$.}
\begin{center}
         {\setlength{\tabcolsep}{0.3cm}
                \renewcommand{\arraystretch}{1.2} 
                 \resizebox{0.9\textwidth}{!}{%
\begin{tabular}{c|ccc|cccc|cc}
\hline
    & \multicolumn{3}{c}{\small$\langle V \rangle$}  & \multicolumn{4}{c}{\small $\langle  r_{eN}^2 \rangle$} &  \multicolumn{2}{c}{\small $\langle  r_{12} \rangle$}  \\[5pt]
Z &     $^{(a)}$ &         $^{(b)}$   & ${}^\text{\cite{Barrois:1996}}$
                                                                                 &     $^{(a)}$   &         $^{(b)}$    &  $^{(g.s.)}$  & $^{(g.s.) \dagger,\ddagger} $   &     $^{(a)}$   &         $^{(b)}$    \\ \hline
3  &    -10.4225  &    -10.4108       &      -10.4255  & 25.591         &        25.729         & 6.1561  & ${}^\dagger$6.1182  & 6.3396  &   6.3464  \\
4  &    -19.2312  &   -19.2139        &     -19.2397   & 10.317         &        10.291         & 2.1877  & ${}^\ddagger$2.1693 & 4.0501 &    4.0483 \\
5  &    -30.7637  &   -30.7765        &     -30.7790   &  5.5628        &          5.5626        &1.1635  & ${}^\ddagger$1.1327 & 2.9815  &   2.9826  \\
10 &   -129.2273  &  -129.2768     &      -129.3181 &  1.0630        &         1.0568       & 0.6207  &  & 1.3053   &   1.3020  \\
15 &   -296.1198  &   -295.8014    &                       &  0.4287        &          0.4320       & 0.2506 &   &  0.8294  &   0.8325 \\
20 &   -530.8696  &   -530.733      &                       &  0.2316        &          0.2319       & 0.1352 &   & 0.6095   &  0.6099 \\
\hline
\end{tabular}}
}
\end{center}
\end{table}

\subsection{Results}

The results of our variational calculations for energies and normalized nucleus-electron cusp parameters for Li-like ions in the spin-quartet state $1^40^+$ (${1}^4S$) are presented in Table \ref{resultsli}. Presented numbers are obtained by using the following wavefunctions:  for $2 < Z \leq 20$ a 9-parametric Ansatz (a) with seed function $\varphi_1$ given by (\ref{Li-psi-1}) and for $2 \leq Z \leq 20$ a 15-parametric Ansatz (b) with seed function $\varphi_3$ given by (\ref{Li-psi-3}). Both trial functions describe 2-3 s.d. correctly in the domain of applicability of the Quantum Mechanics of Coulomb Charges (QMCC). The improvement in the energy obtained with the 15-parameter trial function (b) with respect to the values obtained with the 9-parameter trial function (a) occurs in the 4-5 s.d.  A comparison of the present results with those obtained by other methods in \cite{King}, \cite{Barrois:1996}, \cite{QMC:2014}, \cite{NIST}, shows that the compact trial function (\ref{Li-psi-3}) yields energies that are systematically below the energies obtained by the Hartree-Fock method and reproduce 2-3 s.d. of the most accurate energies.  Among other calculations, Table~\ref{resultsli} shows the results of the Hylleraas CI calculations in \cite{Barrois:1996}  obtained with a basis of dimension(length) 956, the extrapolated Full CI energies calculated in \cite{QMC:2014}, which include double and triple excitations with the basis sets cc-pCVxZ, and (non-variational) fixed-node Diffusion Monte Carlo calculations performed in  \cite{QMC:2014} with a two-configuration Slater-Jastrow form wave function. Full CI results are known to be accurate to about $1\times 10^{-3}\,$a.u. and sometimes they are referred as {\it exact} (see  \cite{QMC:2014}).  For the special case of the Lithium atom, $Z=3$, Table II in \cite{King} presents an extended list of non-relativistic energies for various low-lying spin quartet states. In particular, for the $1^4\,0^+$ state, the most accurate result at present is marked by ${}^{\dagger}$ in Table \ref{resultsli}.

We also calculated the expectation values of the potential $\langle V\rangle$ and $\langle r_{eN}^2 \rangle = \frac{1}{3}\langle \sum_{i=1}^3 r_i^2 \rangle$, i.e. the average electron-nucleus distance squared {\it vs} $Z$ for the spin-quartet state $1^40^+$ (${1}^4S$). In  Table \ref{table2} we show these expectation values obtained with the trial function (\ref{EStrialfunct}) with seed function $\varphi_1$  (\ref{Li-psi-1})  $^{(a)}$ and  with seed function $\varphi_3$  (\ref{Li-psi-3}) $^{(b)}$. Results for $\langle V\rangle$ indicate a very good agreement in 3-4 s.d. between values obtained with the trial function  (\ref{EStrialfunct}) and the values obtained with the large expansions in
Ref.\cite{Barrois:1996}. It is also an indication that our trial functions are uniformly accurate and describe expectation values, other than the energy, with  similarly high accuracy. For $\langle r_{eN}^2 \rangle$ it was also calculated for the spin-quartet state $1^4\,0^+$ for different values of the nuclear charge $Z$ with the trial function (\ref{EStrialfunct}) with seed function $\varphi_1$ (\ref{Li-psi-1}) $^{(a)}$, and with seed function $\varphi_3$ (\ref{Li-psi-3}) $^{(b)}$. There are no results known for this expectation value for the $1^4\,0^+$ to make a comparison with.
Nonetheless, we can get an idea of the accuracy of the expectation value by computing the  expectation value $\langle r_{eN}^2 \rangle$  for the ground state  $1^2 0^+$, in particular   using the Ansatz (d) of the trial function (\ref{GStrialfunct}) and comparing it with the existing results using large multiple basis set in Hylleraas coordinates \cite{Yan:1995},  and  using exponentially gaussian correlated factors \cite{Frolov:2014}. These results indicate that our expectation values coincide in 2-3 s.d. with those obtained using large expansion sets.

For this lowest spin quartet state $1^4\,0^+$ of the Li-like sequence, the energy is well fitted by the Majorana formula
\begin{equation}
\label{Li-majorana}
   E_{M}(1^4\,0^+)\ =\  -\,\frac{49 Z^2}{72}\ +\ 0.358849\, Z\ -\ 0.159566\ .
\end{equation}
It provides an accuracy of 2-3 s.d. for $3 \leq Z \leq 5$ and 3-4 s.d. for $Z \geq 15$ with respect to the most accurate results! Energy values obtained with this formula are presented in Table \ref{resultsli} in the last column.

\begin{figure}[htb]
\begin{center}
  \includegraphics[width=0.5\textwidth ,angle=-90]{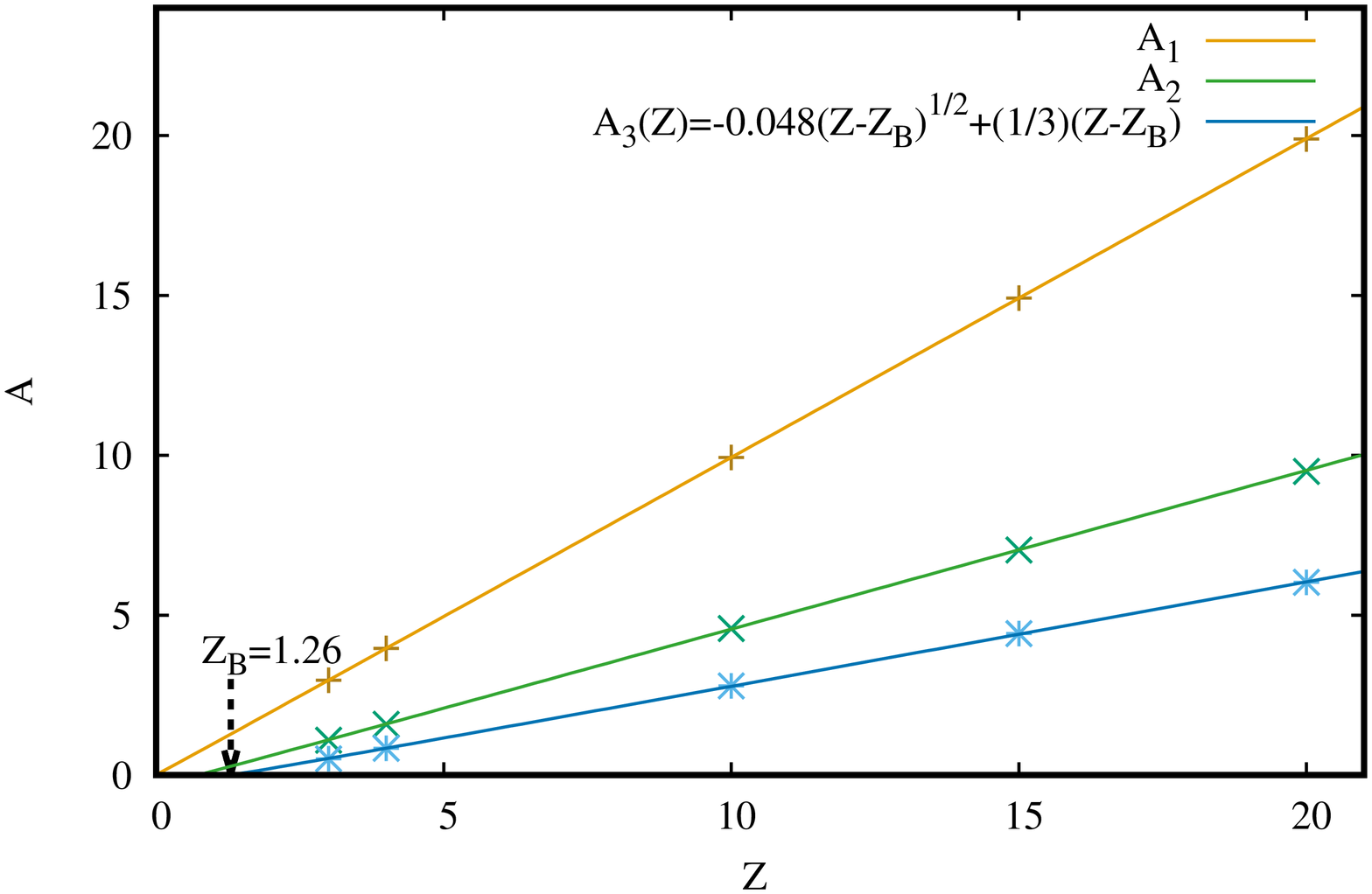}\\
   \includegraphics[width=0.5\textwidth ,angle=-90]{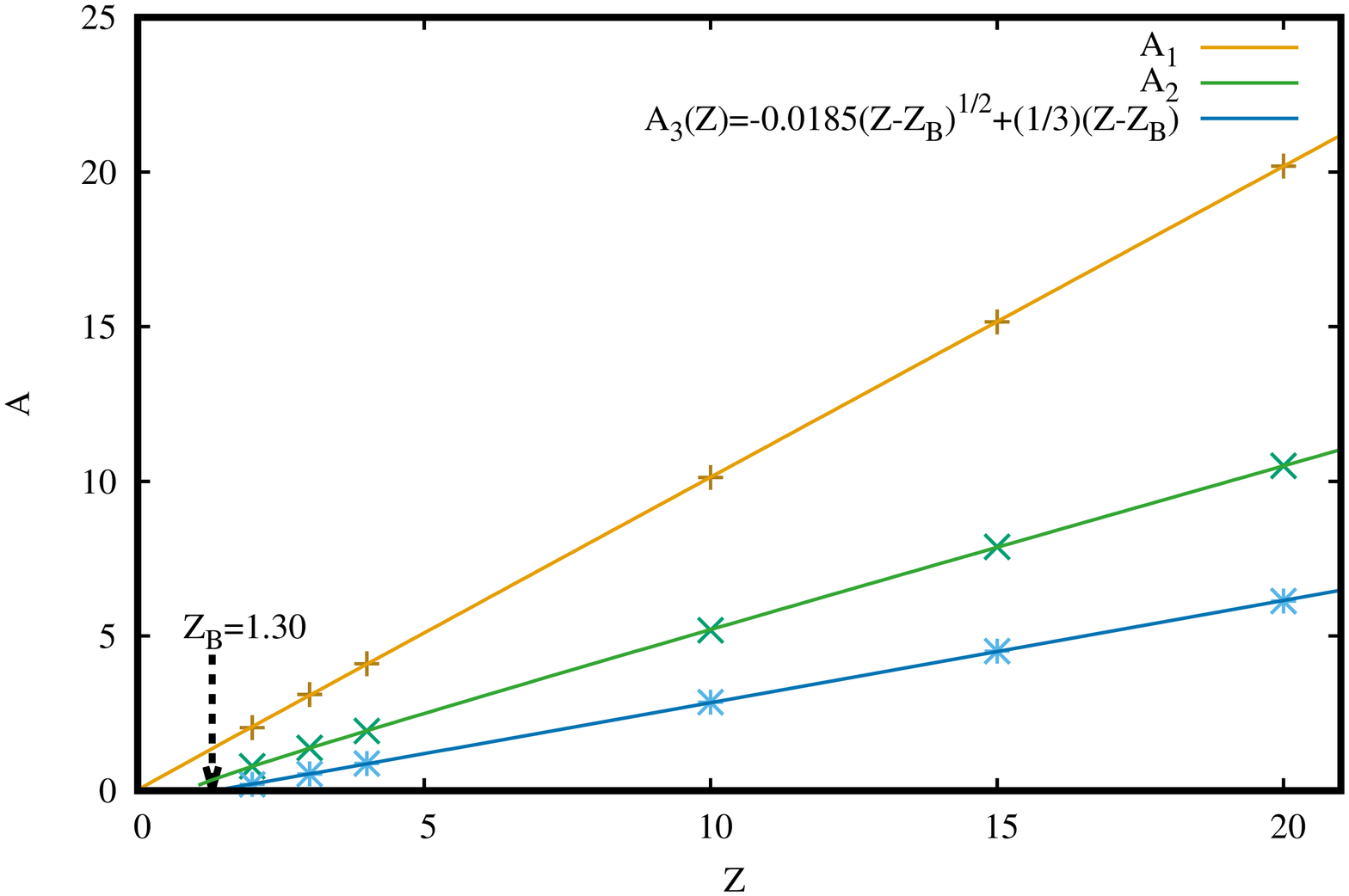}
  \caption{\label{constraintsES}
  Spin-quartet state $1\,{}^4\,0^+$: $A_{1,2,3}^{(\varphi_1,\varphi_2)}$ for Li-like atoms {\it vs} the nuclear charge $Z$:  Ansatz with seed function $\varphi_1$  (top) and with seed function $\varphi_3$  (bottom).  All quantities are fitted by a function of the form (\ref{constraintsf}) with slopes $b=1,1/2,1/3$ for $A_{1,2,3}^{(\varphi_1,\varphi_2)}$, respectively. The overall normalizability is defined by the (smallest) quantity $A_3^{(\varphi_1,\varphi_2)}$ leading to the critical charge $Z_B=1.26$ for $\varphi_1$ and $Z_B=1.3$ for $\varphi_3$.
  }
\end{center}
\end{figure}

%

With regards to the square integrability of the wave function, the quantities $A_{1,2,3}^{(\varphi_1,\varphi_2)}$ for the seed wave function $\varphi_1$ and $\varphi_3$
are plotted in Fig. \ref{constraintsES}. They exhibit an approximate linear behavior as functions of $Z$ and can be fitted by functions of the form (\ref{constraintsf}).    One can observe  from the plots in that figure   that the wave function $\varphi_1$ is not normalizable for $Z\leq Z_B=1.26$ while the wavefunction $\varphi_3$ is not normalizable for $Z \leq 1.30$. In particular, this would mean that for He$^-$ $(Z=2)$ the state   $1\,{}^4\,0^+$   is bound. The variational result for the energy at $Z=2$ is included in Table \ref{resultsli}, even though the spin-quartet ground state energy of ${\rm He}^{-}$  ($E=-2.1656$ a.u.) is larger than that of ${\rm He}$ in $1s\,2s$ state ($E=-2.1752$ a.u.).  Since the corresponding wave function is still normalizable if we follow the behavior of $A_3^{\varphi_3}$   i.e. the state $1\,{}^4\,0^+$ of ${\rm He}^-$ is embedded in the continuum   and lying above the lowest $(1s\,2s\,2p)\ 1^4\,P^-$ state of He$^-$ which has an energy $E=-2.17646$ a.u. \cite{Holoien:1967}. It is not cleat what would happen if a more accurate function for $1\,{}^4\,0^+$  of ${\rm He}^-$  is taken as the trial function. It might be a subject of a separate study.

\section{Conclusions}

In this paper, Part III of the sequence, two (ultra)-compact functions for the Li-like sequence in the spin-quartet state $1^4\,0^+$  are constructed following the principles of physical adequacy formulated and implemented in Part I \cite{Part-1:2020} for the spin-doublet state $1^2\,0^+$ of the Li-like sequence. The simplest function among these functions is a straightforward generalization of the exact (at $Z\to \infty$) wave function by including the inter-electronic correlations in exponential form from one side
and allowing the coefficients in the prefactor of the exact function vary from another side. It contains 9 variational parameters and reproduces 2-3 s.d. in the energy correctly for $Z \leq 20$ as compared to others, probably more accurate results. In the second trial  function the effective Coulomb interactions between electrons and also between nucleus and electrons due to different charge screening at small and large distances was realized additionally. It was done by introducing a simple interpolation function ${\hat r}$ instead of the distance $r$ into the Coulomb orbitals: $\text{exp}(-\al_i {\hat r}_i)$ and $\propto \text{exp}(\al_{jk} {\hat r}_{jk})$, where $\al_i, \al_{jk}$ are parameters. This introduction of physically relevant behavior into the trial function results in a significant improvement in 4-6 s.d. in the variational energy. In the process of optimization it was also observed that the correlation terms $r_{jk}$ in the prefactors of the second trial function does not lead to an improvement (or changes) in the variational energy in the first 4-6 s.d. It implies that these terms can be vanished by removing them from the prefactors without deterioration of the accuracy in the variational energy. Similar situation occurred in Part II in the analysis of the first excited spin-singlet state $2^1 S$ of the He-like sequence, see \cite{Part-2:2021}. Notably, both proposed trial functions describe quite accurately the normalized nucleus-electron cusp with a relative deviation $\sim 0.1\%$ in the domain $3 \leq Z \leq 20$ (see Table \ref{resultsli}), the expectation values of the potential and of electron-nucleus distance squared $\langle r_{eN}^2 \rangle$, which also agree in $\sim 0.1\%$ with other calculations (see Table \ref{table2}).
This analysis indicates that the most accurate trial functions continue to keep, in a first approximation, the functional structure of the prefactor of the exact wave function, which occurs in  the limit $Z\to\infty$. This observation gives a hint how to proceed building the trial functions for other excited states. In particular, for the spin-quartet state $(1s\,2s\,2p_0)$ of the unit total angular momentum $L=1$, which is the state of the lowest energy among the spin-quartet states, see e.g. \cite{Barrois:1996}, a natural proposal for a trial function is to consider seed functions as follows:
\begin{equation}
\label{Li-psi4P}
  \varphi_{{1}^4\,P}\ =\ (1 - a Z r_2)\,(r_3 \cos\theta_3)\,
  e^{-\al_1 Z {\hat r}_1 - \al_2 Z {\hat r}_2 - \al_3 Z {\hat r}_3 + \al_{12} {\hat r}_{12} + \al_{13} {\hat r}_{13} + \al_{23} {\hat r}_{23}}\ .
\end{equation}
Note that this function has an explicit angular dependence. This state will be studied elsewhere.

It is remarkable that the both variational trial functions yield normalized nucleus-electron cusp parameters close to the exact value equal to 1, with relative accuracy $\sim 0.1\%$.
Our variational energies, which might approach to the exact ones from above, are systematically below the energies obtained by the Hartree Fock method, and agree in 2-3 s.d. of the other energies obtained, in particular, in extrapolated Full CI calculations.
Critical charges, at which the used trial functions stop to be normalizable, are found for both states with zero angular momentum, the spin-doublet ground state and for the spin-quartet state, they are $Z_B(1^2\,0^+) \approx 1.65$ and $Z_B(1^4\,0^+) \approx 1.30$, respectively. It might be considered as surprising that the critical charge for the (excited) spin-quartet state $1^4\,0^+$  is significantly smaller than the critical charge for the spin-doublet ground state $1^2\,0^+$(!). The present authors have no explanation for this result.
It must be noted that the variational energies for $Z=2$ obtained in our variational calculations indicate that He$^-$ ion in the spin-quartet state $1^4\, 0^+$ is bound being embedded into the continuum. This result deserves a profound analysis.

\begin{figure}[htb]
\begin{center}
  \includegraphics[width=0.5\textwidth ,angle=0]{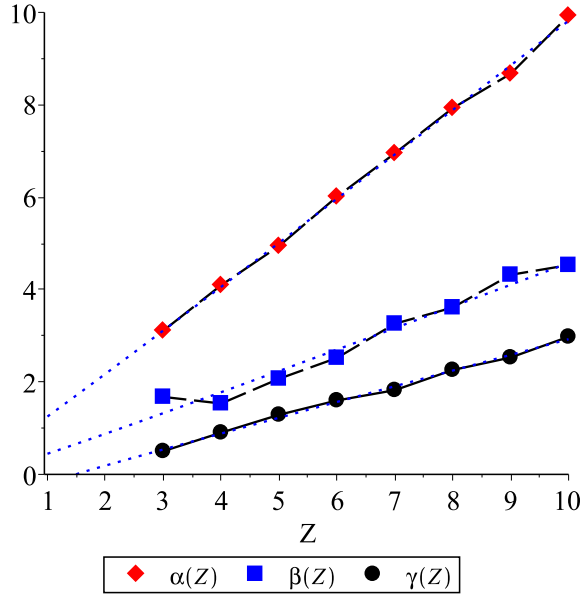}
  \end{center}
  \caption{\label{BarroisParams} Plots of non-linear parameters $\al, \be, \ga$ vs $Z$  for the Hylleraas-CI expansion of the spin-quartet $(1s2s3s)$ wave function (\ref{BarroisPsi})  used in \cite{Barrois:1996}. The dotted lines correspond to fits (\ref{BarroisFits}), which have the correct linear slope at $Z\to\infty$.  The extrapolation of the fit for $\ga(Z)$ to domain where it vanishes $\ga(Z)=0$ yields a critical charge $Z_B \approx 1.4974$ for which the wave function (\ref{BarroisPsi}) stops to be normalizable (see the text).}
\end{figure}

A similar analysis of the square integrability of the spin-quartet state $(1s\,2s\,3s)$ of the lithium iso-electronic series can be done for the Hylleraas-CI trial function
\begin{equation}
\label{BarroisPsi}
  \Psi_{{}^4S}\ =\ \sum_i c_i\ {\cal A} \{r_1^{a_i}r_2^{b_i}r_3^{c_i}r_{23}^{\lambda_i}r_{13}^{\mu_i}r_{12}^{\gamma_i} \,
  \exp[-(\al r_1+\be r_2 +\ga r_3)]\,  \chi \}\ ,
\end{equation}
used in \cite{Barrois:1996}; here ${\cal A}$ is the antisymmetrizer (\ref{Asym}) and $\chi$ is a three-electron (symmetric) eigenfunction of spin $3/2$.
In Fig.\ref{BarroisParams} we have plotted the non-linear parameters $\al,\be,\ga$ {\it vs} $Z$ ($3\leq Z \leq10$) presented in \cite{Barrois:1996} in Table 4.
From these plots it is seen clear that natural smoothness of these parameters {\it vs} $Z$ does not occur, which we think it must, it was not taken into account in \cite{Barrois:1996}. It indicates a deficiency in the optimization procedure used in \cite{Barrois:1996} in finding those parameters.
Although the approximate linear behavior of these parameters {\it vs} $Z$ seems evident (except for $Z=3,4$), a straightforward linear fit shows that the linear slopes of $\al,\be,\ga$, being, respectively, $\sim 0.9584, 0.4675, 0.3405$, fail to reproduce the correct asymptotic behavior as $Z \to \infty$ that should be $1, 1/2, 1/3$, correspondingly.
It is one more indication to the deficiency of the optimization procedure performed in \cite{Barrois:1996}. If linear slops found in \cite{Barrois:1996} remain one can expect that the results obtained in (\ref{BarroisPsi}) will lose accuracy significantly for larger values of the charge $Z > 10$. It will happen not only for the energy, but, more importantly, for expectation values, which require higher accuracy in a domain different from one giving dominant contribution to the energy, especially, when at large distances $r \to \infty$ - this domain is very relevant when dispersion processes are studied.

In spite of the deficiencies in localizing optimal parameters $\al,\be,\ga$\, carried out in \cite{Barrois:1996}, the analysis of square integrability of the function (\ref{BarroisPsi}) can be performed in a complete analogy to one for the parameters $A_{1,2,3}$ introduced beforehand for the compact functions for the states $1^2\, 0^+$ and $1^4\, 0^+$. Fitting $\al,\be,\ga$ with a function (\ref{constraintsf}) while fixing {\it ad hoc} the correct linear slopes at large $Z \to \infty$, we get
\begin{align}
 \al(Z) &\ =\ -0.21592\, \sqrt{Z + 0.51853}\ +\ (Z + 0.51853)\  , \non
\\
 \be(Z) &\ =\ -0.17973\, \sqrt{Z + 0.30397}\ +\ \frac{1}{2}\,(Z + 0.30397)\ ,
\label{BarroisFits}
\\
 \ga(Z) &\ =\ 0.03155\, \sqrt{Z - 1.49737}\ +\ \frac{1}{3}\,(Z - 1.49737)\ , \non
\end{align}
see Fig.\ref{BarroisParams}. Following our previous experience, the fitted parameters (\ref{BarroisFits}) should lead to more accurate energies than ones obtained in \cite{Barrois:1996}.

One can see explicitly in Fig.\ref{BarroisParams} that there exists a hierarchy
$\ga < \be < \al$ for all $Z$ and the normalizability of the trial function depends
mostly on the smallest of these parameters $\ga$.  Therefore, from the  fit of $\ga(Z)$ (\ref{BarroisFits}), we can see that the function (\ref{BarroisPsi}) is characterized by the critical charge at \hbox{$Z_B( 1^4\,0^+) \approx 1.4974$}, where it vanishes, $\ga(Z)=0$, and the function (\ref{BarroisPsi}) stops to be normalizable.
This value of the critical charge is larger than our estimates $Z_B(1^4\,0^+) \sim 1.26 - 1.30$ and, thus, it represents a better approximation to the critical charge. In any case,
for $Z=2$ (${\rm He}^-$) the systems is predicted to be bound in the $1^4\,0^+$ spin-quartet state. We can only assume that similar situation with the non-linear parameters but used for the ground state, or other quartet states like ${}^4\,P$ studied within the Hylleraas-CI expansions must hold. Present authors were unable to find such an information for those states.

In Parts I, II and III it was developed a formalism to construct some compact trial functions
for few states of He-like and Li-like sequences for $Z \leq 20$ leading to reasonably accurate variational energies, expectation values and cusp parameters. The formalism is based on three elements:
\begin{itemize}
  \item employing the exact eigenfunctions at $Z=\infty$, where the problem is reduced to the study of the system of two-three non-interacting, independent Hydrogen atoms, taking the parameters of the Coulomb orbitals as variational,

  \item introducing electronic correlations in the form of $r_{ij}$-dependence in exponential and pre-exponential forms,

  \item introducing electron-electron and electron-nuclear (anti)-screening in the form of meromorphic (rational) function in the exponents.
\end{itemize}
Additionally, it is assumed that variational parameters are smooth functions of the charge $Z$.
Evidently, this formalism can be extended to other excited states of the He-like and Li-like sequences as well as to other atomic systems like Be-like sequence. It will be done elsewhere.

\section*{Acknowledgments}

\noindent
D.J.N. is supported in part by PRODEP project 42027 UV-CA-320 (Mexico). J.C.~del\,V. is supported by CONACyT PhD Grant No.570617 (Mexico) in early stage of the work and by postdoctoral grant via DGAPA grant IN113819 (Mexico) and SNI assistance grant.
This work is partially supported by CONACyT grant A1-S-17364 and DGAPA grants IN113819, IN113022 (Mexico). A.V.T. thanks PASPA-UNAM for a support during his sabbatical stay at University of Miami, where this work was finished.

\end{document}